\documentclass{article}

\usepackage{amsfonts}
\usepackage{amsmath}
\usepackage[english]{babel}
\usepackage{graphicx}
\usepackage{graphics}
\usepackage{epsfig}
\input epsf

\newtheorem{proposition}{Proposition}
\newtheorem{lemma}{Lemma}

\newcommand{\eps}{\varepsilon}

\newcommand{\one}{{\bf 1}}
\newcommand{\zero}{{\bf 0}}
\newcommand{\pidn}{\pi_{\mbox{\tiny DN}}}
\newcommand{\onedn}{{\bf 1}}
\newcommand{\piscc}{\pi_{\mbox{\tiny IN+SCC}}}
\newcommand{\out}{{\mbox{\tiny OUT}}}
\newcommand{\scc}{{\mbox{\tiny IN+SCC}}}
\newcommand{\escc}{{\mbox{\tiny ESCC}}}
\newcommand{\pureout}{{\mbox{\tiny PureOUT}}}
\newcommand{\onesccn}{{\bf 1}}
\newcommand{\onescc}{{\bf 1}}
\newcommand{\piout}{\pi_{\mbox{\tiny OUT}}}
\newcommand{\oneout}{{\bf 1}}

\def\u{{{\bf u}_{\mbox{\tiny IN+SCC}}}}
\def\v{{\bf v}}
\def\ue{{{\bf u}_{\mbox{\tiny ESCC}}}}

\begin{document}

\title{Distribution of PageRank Mass Among\\ Principle Components of the Web}

\author{Konstantin Avrachenkov\thanks{ 
INRIA Sophia Antipolis, 2004, Route des Lucioles,
06902, France, E-mail: k.avrachenkov@sophia.inria.fr}
\and Nelly Litvak\thanks{ 
University of Twente, Dept. of Applied Mathematics, P.O.~Box~217,
7500AE~Enschede, The Netherlands,
E-mail: n.litvak@ewi.utwente.nl}
\and Kim Son Pham\thanks{
St.Petersburg State
University,35, University Prospect, 198504, Peterhof, St.Petersburg, Russia,
E-mail: sonsecure@yahoo.com.sg }}

\date{September 2007}

\maketitle

\begin{abstract}
We study the PageRank mass of principal components in a bow-tie Web
Graph, as a function of the damping factor~$c$. Using a singular
perturbation approach, we show that the PageRank share of IN and SCC
components remains high even for very large values of the damping
factor, in spite of the fact that it drops to zero when $c\to 1$.
However, a detailed study of the OUT component reveals the presence
``dead-ends'' (small groups of pages linking only to each other)
that receive an unfairly high ranking when $c$ is close to one. We
argue that this problem can be mitigated by choosing $c$ as small as
1/2.
\end{abstract}

%
%

\section{Introduction}

The link-based ranking schemes such as PageRank~\cite{Page98},
HITS~\cite{Kleinberg99}, and SALSA~\cite{Lempel00} have been
successfully used in search engines to provide adequate importance
measures for Web pages. In the present work we restrict ourselves to
the analysis of the PageRank criterion and use the following
definition of PageRank from~\cite{Langville03}. Denote by $n$ the
total number of pages on the Web and define the $n\times n$
hyper-link matrix $W$ as follows:
\begin{equation}
\label{eq:w} w_{ij} = \left\{ \begin{array}{ll}
1/d_i, & \mbox{if page $i$ links to $j$},\\
1/n, & \mbox{if page $i$ is dangling},\\
0, & \mbox{otherwise},
\end{array} \right.
\end{equation}
for $i,j=1,...,n$, where $d_i$ is the number of outgoing links from
page $i$. A page is called {\it dangling} if it does not have
outgoing links. The PageRank is defined as a stationary distribution
of a Markov chain whose state space is the set of all Web pages, and
the transition matrix is
\begin{equation}
\label{GoogleMatrix} G = cW + (1-c)(1/n)\one^T\one.
\end{equation}
Here and throughout the paper we use the
symbol $\one$ for a column vector of ones having by default an
appropriate dimension. In (\ref{GoogleMatrix}), $\one^T\one$ is a matrix whose all entries
are equal to one, and $c \in (0,1)$ is the parameter known as a {\it damping factor}. Let $\pi$ be
the PageRank vector. Then by definition, $ \pi G = \pi$, and
$||\pi||=\pi\one =1$, where we write $||{\bf x}||$ for the
$L_1$-norm of vector ${\bf x}$.

The damping factor $c$ is a crucial parameter in the PageRank
definition. It regulates the level of the uniform noise introduced
to the system. Based on the publicly available information Google
originally used $c=0.85$, which appears to be a reasonable
compromise between the true reflection of the Web structure and
numerical efficiency (see \cite{LangvilleMeyer} for more detail).
However, it was mentioned in \cite{Boldi05} that the value of $c$
too close to one results into distorted ranking of important pages.
This phenomenon was also independently observed
in~\cite{Avrachenkov06SM}. Moreover, with smaller $c$, the PageRank
is more rebust, that is, one can bound the influence of outgoing
links of a page (or a small group of pages) on the PageRank of other
groups~\cite{Bianchini05} and on its own
PageRank~\cite{Avrachenkov06SM}.

In this paper we explore the idea of relating the choice of $c$ to
specific properties of the Web structure. In papers
\cite{Broder00,Kumar00} the authors have shown that the Web graph
can be divided into three principle components. The Giant Strongly
Connected Component (SCC) contains a large group of pages all having
a hyper-link path to each other. The pages in the IN (OUT) component
have a path to (from) the SCC, but not back. Furthermore, the SCC
component is larger than the second largest strongly connected
component by several orders of magnitude.

In
Section~\ref{sec:ergodic} we consider a Markov walk governed by the
hyperlink matrix~$W$ and explicitly describe the limiting behavior
of the PageRank vector as $c\to 1$. We experimentally study the
OUT component in more detail to discover a so-called Pure OUT
component (the OUT component without dangling nodes and their
predecessors) and show that Pure OUT contains a number of small
sub-SCC's, or dead-ends, that absorb the total PageRank mass when
$c=1$. In Section~\ref{sec:scc} we apply the singular perturbation
theory~\cite{Avrachenkov99thesis,KorolyukTurbin,PervozvanskiiGaitsgori,YinZhang}
to analyze the shape of the PageRank of IN+SCC as a function of
$c$. The dangling nodes turn out to play an unexpectedly important
role in the qualitative behavior of this function. Our analytical
and experimental results suggest that the PageRank mass of IN+SCC
is sustained on a high level for quite large values of
$c$, in spite of the fact that it drops to zero as $c\to 1$.
 Further, in Section~\ref{sec:bounds} we show that the total
PageRank mass of Pure OUT component increases with $c$. We argue
that $c=0.85$ results in an inadequately high ranking for Pure OUT
pages and we present an argument for choosing $c$ as small as 1/2.
We confirm our theoretical argument by experiments with log files.
We would like to mention that the value $c=1/2$ was also used in
\cite{PRcitations} to find gems in scientific citations. This choice
was justified intuitively by stating that researchers may check
references in cited papers but on average they hardly go deeper than
two levels. Nowadays, when search engines work really fast, this
argument also applies to Web search. Indeed, it is easier for the
user to refine a query and receive a proper page in fraction of
seconds than to look for this page by clicking on hyper-links.
Therefore, we may assume that a surfer searching for a page, on
average, does not go deeper than two clicks.

The body of the paper contains main ideas and results. The
necessary information from the perturbation theory and the proofs
are given in Appendix.

\section{Datasets}
\label{sec:datasets}

We have collected two Web graphs, which we denote by INRIA and
FrMathInfo. The Web graph INRIA was taken from the site of INRIA,
the French Research Institute of Informatics and Automatics. The
seed for the INRIA collection was Web page {\tt www.inria.fr}. It is
a typical large Web site with around 300.000 pages and 2 millions
hyper-links. We have collected all pages belonging to INRIA. The Web
graph FrMathInfo was crawled with the initial seeds of 50
mathematics and informatics laboratories of France, taken from
Google Directory. The crawl was executed by Breadth First Search of
depth~6. The FrMathInfo Web graph contains around 700.000 pages and
8 millions hyper-links. Because of the fractal structure of the Web
\cite{Dill02} we expect our datasets to be enough representative.

The link structure of the two Web graphs is stored in Oracle
database. We could store the adjacency lists in RAM to speed up the
computation of PageRank and other quantities of interest. This
enables us to make more iterations, which is extremely important
when the damping factor $c$ is close to one. Our PageRank
computation program consumes about one hour to make 500 iterations
for the FrMathInfo dataset and about half an hour for the INRIA
dataset for the same number of iterations. Our algorithms for
discovering the structures of the Web graph are based on Breadth
First Search and Depth First Search methods, which are linear in the
sum of number of nodes and links.

\section{The structure of the hyper-link transition matrix}
\label{sec:ergodic}

With the bow-tie Web structure~\cite{Broder00,Kumar00} in mind, we
would like to analyze a stationary distribution of a Markov random
walk governed by the hyper-link transition matrix $W$ given by
(\ref{eq:w}). Such random walk follows an outgoing link chosen
uniformly at random, and dangling nodes are assumed to have links to
all pages in the Web. We note that the methods presented below can
be easily extended to the case of personalized
PageRank~\cite{Haveliwala03}, when after a visit to a dangling node,
the next page is sampled from some prescribed distribution.

Obviously, the graph induced by $W$ has a much higher connectivity
than the original Web graph. In particular, if the random walk can
move from a dangling node to an arbitrary node with the uniform
distribution, then the Giant SCC component increases further in
size. We refer to this new strongly connected component as the
Extended Strongly Connected Component (ESCC). Due to the artificial
links from the dangling nodes, the SCC component and IN component
are now inter-connected and are parts of the ESCC. Furthermore, if
there are dangling nodes in the OUT component, then these nodes
together with all their predecessors become a part of the ESCC.

\begin{figure}[h]
  \begin{minipage}{0.4\textwidth}
    \begin{center}
        \centering {\epsfxsize=2.0in \epsfbox{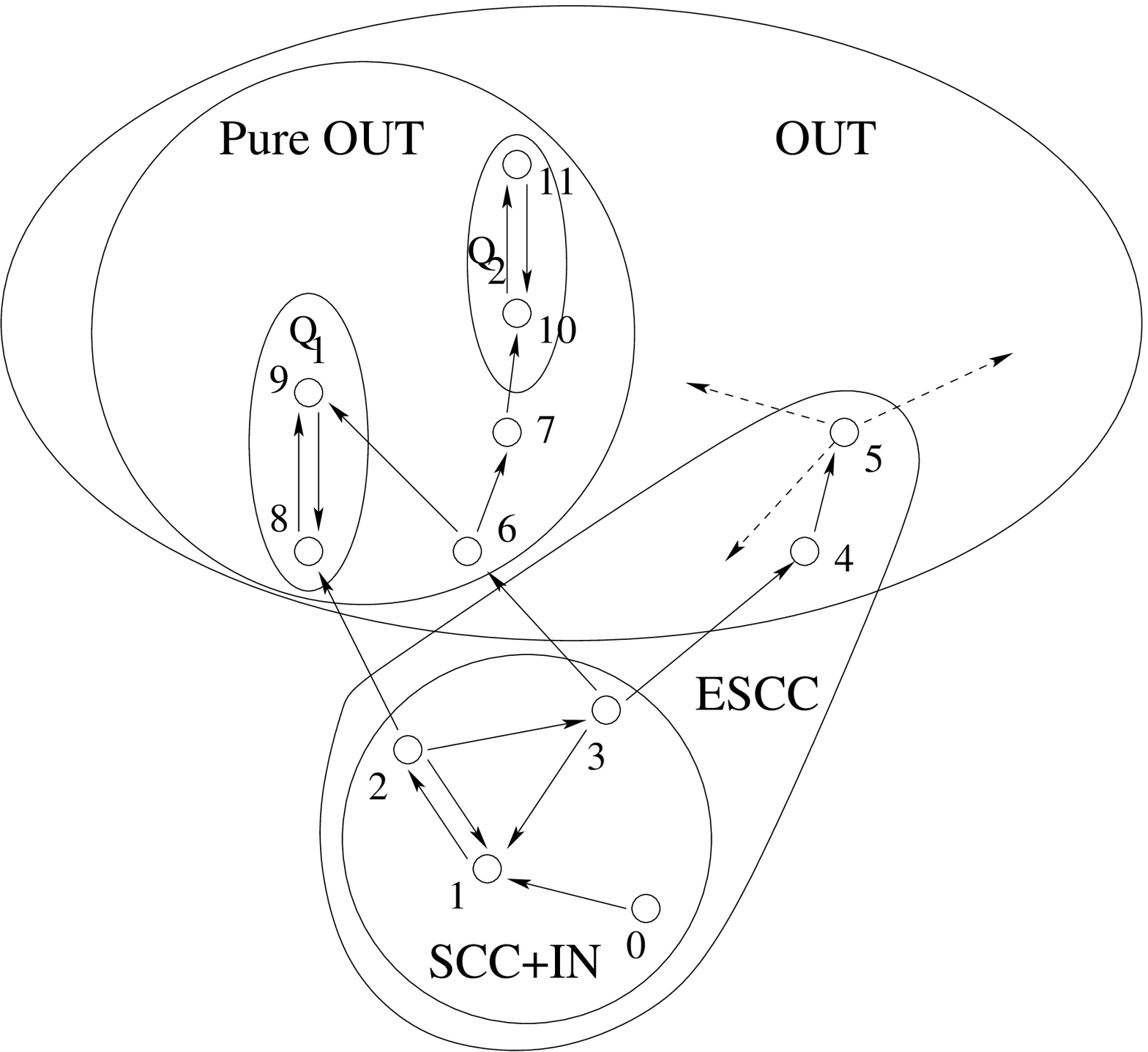}}
        \caption{Example of a graph}
        \label{fig:smallgraph}
    \end{center}
  \end{minipage}
\nolinebreak
  \begin{minipage}{0.4\textwidth}
    \begin{center}
\centering {\begin{tabular}{|r|r|r|} \hline
\# &$INRIA$&$FrMathInfo$\\
\hline \hline
total nodes & 318585 & 764119\\
nodes in SCC  & 154142 & 333175\\
nodes in IN  & 0 & 0 \\
nodes in OUT  & 164443 & 430944 \\
nodes in ESCC & 300682 & 760016\\
nodes in Pure OUT & 17903 & 4103\\
SCCs in OUT & 1148 & 1382\\
SCCs in Pure Out & 631 & 379\\
\hline
\end{tabular}} \caption{Component sizes in INRIA and FrMathInfo
datasets} \label{tab:sizes}
    \end{center}
  \end{minipage}
\end{figure}

In the mini-example in Figure~\ref{fig:smallgraph},
node~0 represents the IN component, nodes from 1 to 3 form the SCC
component, and the rest of the nodes, nodes from 4 to 11, are in
the OUT component. Node~5 is a dangling node, thus, artificial
links go from the dangling node~5 to all other nodes. After
addition of the artificial links, all nodes from 0 to 5 form the ESCC.

%
%

In the Markov chain induced by the matrix $W$, all states from ESCC
are {\it transient}, that is, with probability~1, the Markov chain
eventually leaves this set of states and never returns back. The
stationary probability of all these states is zero. The part of the
OUT component without dangling nodes and their predecessors forms a
block that we refer to as a Pure OUT component. In
Figure~\ref{fig:smallgraph} the Pure OUT component consists of nodes
from 6 to 11. Typically, the Pure OUT component is much smaller than
the Extended SCC. However, this is the set where the total
stationary probability mass is concentrated. The sizes of all
components for our two datasets are given in Figure~\ref{tab:sizes}.
Here the size of the IN components is zero because in the Web
crawl we used the Breadth First Search method and we started from
important pages in the Giant SCC. For the purposes of the
present research it does not make any difference since we always
consider IN and SCC together.

Let us now analyze the structure of the Pure OUT component in more
detail. It turns out that inside  Pure OUT there are many disjoint
strongly connected components. All states in these sub-SCC's (or,
``dead-ends'') are {\it recurrent}, that is, the Markov chain
started from any of these states always returns back to it. In
particular, we have observed that there are many dead-ends of size 2
and 3.
%
%
%
The Pure OUT component also contains transient states that
eventually bring the random walk into one of the dead-ends. For
simplicity, we add these states to the giant transient ESCC
component.

Now, by appropriate renumbering of the states, we can refine the
matrix $W$ by subdividing all states into one giant transient block
and a number of small recurrent blocks as follows:
 \begin{equation}
\label{eq:ESCCdetail} W=\left[ \begin{array}{cccc}
Q_1 &        & 0   & 0 \\
    & \ddots &     &   \\
0   &        & Q_m & 0 \\
{R}_1 & \cdots & {R}_m & {T}
\end{array} \right]\;
\begin{array}{l}\mbox{dead-end (recurrent)}\\
\cdots\phantom{\ddots}\\
\mbox{dead-end (recurrent)}\\
\mbox{ESCC+[transient states in Pure OUT] (transient)}\end{array}
\end{equation}
Here for $i=1,\ldots,m$, a block $Q_i$  corresponds to transitions
inside the $i$-th recurrent block, and a block $R_i$ contains
transition probabilities from transient states to the $i$-th
recurrent block. Block ${T}$ corresponds to transitions between the
transient states. For instance, in example of the graph from
Figure~\ref{fig:smallgraph}, the nodes 8 and 9 correspond to block
$Q_1$, nodes 10 and 11 correspond to block $Q_2$, and all other
nodes belong to block $T$.

We would like to emphasis that the recurrent blocks here are really
small, constituting altogether about 5\% for INRIA and about 0.5\%
for FrMathInfo. We believe that for larger data sets, this
percentage will be even less. By far most important part of the
pages is contained in the ESCC, which constitutes the major part of
the giant transient block.

Next, we note that if $c<1$, then all states in the Markov chain
induced by the Google matrix $G$ are recurrent, which automatically
implies that they all have positive stationary probabilities.
However, if $c=1$, the majority of pages turn into transient states
with stationary probability zero. Hence, the random walk governed by
the Google transition matrix (\ref{GoogleMatrix}) is in fact a
singularly perturbed Markov chain. Informally, by singular
perturbation we mean relatively small changes in elements of the
matrix, that lead to altered connectivity and stationary behavior of
the chain. Using the results of the singular perturbation theory
(see e.g.,
\cite{Avrachenkov99thesis,KorolyukTurbin,PervozvanskiiGaitsgori,YinZhang}),
in the next proposition we characterize explicitly the limiting
PageRank vector as $c\to 1$ (see Appendix~\ref{app:proofs} for the
proof).

\begin{proposition}
\label{prop:pure_out}
Let ${\pi}_{\out,i}$ be a stationary distribution of the Markov
chain governed by $Q_i$, $i=1,\ldots,m$.
Then, we have
\[
\lim_{c \to 1} \pi(c)= \left[{\pi}_{\out,1} \ \cdots \ {\pi}_{\out,m} \
\zero \right],
\]
where
\begin{equation}
\label{eq:barpi} {\pi}_{\out,i}=
\left(\frac{\mbox{\# nodes in block $Q_i$}}{n}+\frac{1}{n}\one^T[I-{T}]^{-1}{R}_i\one\right)\bar{\pi}_{\out,i}
\end{equation}
for $i=1,...,m$, $I$ is the identity matrix, and $\zero$ is a row
vector of zeros that correspond to stationary probabilities of the
states in the transient block.
\end{proposition}

The second term inside the brackets in formula (\ref{eq:barpi})
corresponds to the Page\-Rank mass received by a dead-end
 from the Extended SCC. If $c$ is close to one, then this
contribution can outweight by far the fair share of the PageRank,
whereas the PageRank mass of the giant transient block decreases to
zero. How large is the neighborhood of one where the ranking is
skewed towards the Pure OUT? Is the value $c=0.85$ already too
large? We will address these questions in the remainder of the
paper. In the next section we analyze the PageRank mass IN+SCC
component, which is an important part of the transient block.

\section{PageRank mass of IN+SCC}
\label{sec:scc}

In Figure~3 we depict the PageRank mass of the giant component
IN+SCC, as a function of the damping factor, for FrMathInfo.
\begin{figure}
\centering
\includegraphics[width=6cm]{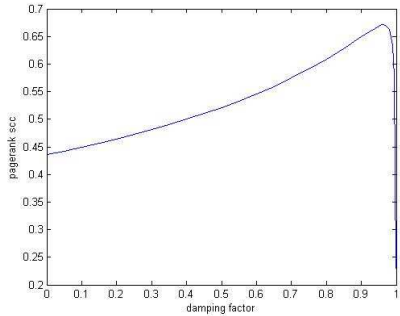} \label{fig:in+scc}
\caption{The PageRank mass of IN+SCC as a function of $c$.}
\end{figure}
Here we see a typical behavior also observed for several pages in
the mini-web from~\cite{Boldi05}: the PageRank first grows with $c$
and then decreases to zero. In our case, the PageRank mass of IN+SCC
drops drastically starting from some value $c$ close to one. We can
explain this phenomenon by highlighting the role of the dangling
nodes.

We start the analysis by subdividing the Web graph sample into three
subsets of nodes: IN+SCC, OUT, and the set of dangling nodes DN. We
assume that no dangling node originates from OUT. This simplifies
the derivation but does not change our conclusions. Then the Web
hyper-link matrix $W$ in (\ref{eq:w}) can be written in the form
\[
W = \left[
\begin{array}{ccc}
Q & 0 & 0 \\
R & P & S \\
\frac{1}{n}\onedn\oneout^T & \frac{1}{n}\onedn\onescc^T &
\frac{1}{n}\onedn\onedn^T
\end{array} \right]\;
\begin{array}{l}\mbox{OUT}\\\mbox{IN+SCC\;\;,}\\\mbox{DN}
\end{array}
\]
where the block $Q$ corresponds to the hyper-links inside the OUT
component, the block $R$ corresponds to the hyper-links from IN+SCC
to OUT, the block $P$ corresponds to the hyper-links inside the
IN+SCC component, and the block $S$ corresponds to the hyper-links
from SCC to dangling nodes.  In the above, $n$ is the total number
of pages in the Web graph sample, and the blocks $\one\one^T$ are
the matrices of ones adjusted to appropriate dimensions.

Dividing the PageRank vector in segments corresponding to the
blocks OUT, IN+SCC and DN,
\[ \pi=[\piout \, \piscc \, \pidn],\]
we can rewrite the well-known formula (see e.g.~\cite{Moler})
\begin{equation}
\label{eq:vector_eq} \pi=\frac{1-c}{n} \one^T [I-cP]^{-1}
\end{equation}
as a system of three linear equations:
\begin{align}
&\label{eq:out} \piout [I-cQ] -\piscc cR
-\frac{c}{n}\pidn\onedn\oneout^T = \frac{1-c}{n}\oneout^T,
\\
&\label{eq:scc} \piscc [I-cP]
-\frac{c}{n}\pidn\onedn\onescc^T = \frac{1-c}{n}\onescc^T,\\
&\label{eq:dn} -\piscc cS +\pidn -\frac{c}{n}\pidn\onedn\onedn^T =
\frac{1-c}{n}\onedn^T.
\end{align}
Solving (\ref{eq:out}--\ref{eq:dn}) for $\piscc$ we obtain
\begin{equation}
\label{eq:piscc} \piscc(c) =
\frac{(1-c)\alpha}{1-c\beta}\u\left[I-cP-\frac{c^2\alpha}{1-c\beta}S\onedn
\u\right]^{-1},
\end{equation}
where \[\alpha= |IN+SCC|/{n}\quad\mbox{and}\quad \beta=|DN|/{n}\]
are the fractions of nodes in IN+SCC and DN, respectively, and
$\u=|IN+SCC|^{-1}\onescc^T$ is a uniform probability row-vector of
dimension $|IN+SCC|$. The detailed derivation of (\ref{eq:piscc})
can be found in Appendix~\ref{app:proofs}.

Now, define
\begin{equation}
\label{eq:kU} k(c)=\frac{(1-c)\alpha}{1-c\beta}, \quad \mbox{and}
\quad U(c)=P+\frac{c\alpha}{1-c\beta}S\onedn \u.
\end{equation}
Then the derivative of $\piscc(c)$ with respect to $c$ is given by
\begin{eqnarray}
\label{piprime} \pi'_\scc(c) = \u\left\{k'(c)I +
k(c)[I-cU(c)]^{-1}(cU(c))'\right\}[I-cU(c)]^{-1},
\end{eqnarray}
where using (\ref{eq:kU}) after simple calculations we get
\[
k'(c)=-\frac{(1-\beta)\alpha}{(1-c\beta)^2},\quad
(cU(c))'=U(c)+\frac{c\alpha}{(1-c\beta)^2}S\onedn \u.
\]
Let us consider the point $c=0$. Using (\ref{piprime}), we obtain
\begin{eqnarray}
\label{eq:piscc'(0)}
\pi'_\scc(0) & = & -\alpha(1-\beta)\u +\alpha\u P.
\end{eqnarray}
One can see from the above equation that the PageRank of pages in
IN+SCC with many incoming links will increase as $c$ increases from
zero, which explains the graphs presented in \cite{Boldi05}.

Next, let us analyze the total mass of the IN+SCC component. From
(\ref{eq:piscc'(0)}) we obtain
$$
||\pi'_\scc(0)||=-\alpha(1-\beta)\u +\alpha\u
P\onescc=\alpha(-1+\beta+p_1),
$$
where $p_1=\u P \onescc$ is the probability that a random walk on
the hyperlink matrix stays in IN+SCC for one step if the initial
distribution is uniform over IN+SCC.  If $1-\beta<p_1$ then the
derivative at 0 is positive. Since dangling nodes typically
constitute more than 25\% of the graph~\cite{Eiron04}, and $p_1$
is usually close to one, the condition $1-\beta<p_1$ seems to be
comfortably satisfied in Web samples. Thus, the total PageRank
of the IN+SCC increases in $c$ when $c$ is small. Note by the way
that if $\beta=0$ then $||\piscc(c)||$ is strictly decreasing in $c$.
Hence, surprisingly, the presence of dangling
nodes qualitatively changes the behavior of the IN+SCC PageRank
mass.

Now let us consider the point $c=1$. Again using (\ref{piprime}),
we obtain
\begin{equation}
\label{piprimec1} \pi'_\scc(1) = -
\frac{\alpha}{1-\beta} \u
[I-P-\frac{\alpha}{1-\beta}S\onedn\u]^{-1}.
\end{equation}
Note that the matrix in the square braces is close to singular.
Denote by $\bar{P}$ the hyper-link matrix of IN+SCC when the outer
links are neglected. Then, $\bar{P}$ is an irreducible stochastic
matrix. Denote its stationary distribution by $\bar{\pi}_{\scc}$.
Then we can apply Lemma~\ref{lm:sp} from the singular perturbation
theory to (\ref{piprimec1}) by taking
\[
A = \bar{P},\quad
\eps C = \bar{P} - P - \frac{\alpha}{1-\beta}S\onedn\u,
\]
and noting that $ \eps C \onescc = R\oneout
+(1-\alpha-\beta)(1-\beta)^{-1}S\onedn. $ Combining all terms
together and using $\bar{\pi}_\scc \onescc = ||\bar{\pi}_\scc||=1$
and $\u \onescc = ||\u||=1$, from (\ref{eq:LaurentX}) we obtain
\begin{align*}
||\pi'_\scc(1)|| &\approx
- \frac{\alpha}{1-\beta} \frac{1}{\bar{\pi}_\scc R\oneout
+\frac{1-\beta-\alpha}{1-\beta}\bar{\pi}_\scc S\onedn}.
\end{align*}
It is expected that the value of $\bar{\pi}_\scc R\oneout
+\frac{1-\beta-\alpha}{1-\beta}\bar{\pi}_\scc S\onedn$ is typically
small (indeed, in our dataset $INRIA$, the value is 0.022), and
hence the mass\\  $||\piscc(c)||$ decreases very fast as $c$
approaches one.

Having described the behavior of the PageRank mass $||\piscc(c)||$
at the boundary points $c=0$ and $c=1$, now we would like to show
that there is at most one extremum on $(0,1)$. It is sufficient to
prove that if $||\piscc'(c_0)||\le 0$ for some $c_0\in(0,1)$ then
$||\piscc'(c)||\le 0$ for all $c>c_0$. To this end, we apply the
Sherman-Morrison formula to (\ref{eq:piscc}), which yields
\begin{equation}
\label{eq:sh-m2}
\piscc(c)=\tilde{\pi}_\scc(c)+\frac{\frac{c^2\alpha}{1-c\beta}
\u[I-cP]^{-1}S\onedn}{1+\frac{c^2\alpha}{1-c\beta}\u[I-cP]^{-1}S\onedn}\,\tilde{\pi}_\scc(c),
\end{equation}
where
\begin{equation}
\label{eq:pitilda}
\tilde{\pi}_{\scc}(c)=\frac{(1-c)\alpha}{1-c\beta}\u[I-cP]^{-1}.\end{equation}
represents the main term in the right-hand side of (\ref{eq:sh-m2}).
(The second summand in (\ref{eq:sh-m2}) is about 10\% of the total
sum for the $INRIA$ dataset for $c=0.85$.) Now the behavior of
$\pi_\scc(c)$ in Figure~3 can be explained by means of the next
proposition (see Appendix~\ref{app:proofs} for the proof).
\begin{proposition}
\label{prop:pitilda} The term $||\tilde{\pi}_\scc(c)||$ given by
(\ref{eq:pitilda}) has exactly one local maximum at some $c_0\in
[0,1]$. Moreover, $||\tilde{\pi}''_\scc(c)||<0$ for $c\in
(c_0,1]$.
\end{proposition}
We conclude that
 $||\tilde{\pi}_\scc(c)||$ is decreasing and concave
for $c\in[c_0,1]$, where $||\tilde{\pi}'_\scc (c_0)||=0$. This is exactly the behavior we observe in the
experiments. The analysis and experiments suggest that $c_0$ is definitely larger than 0.85 and actually is quite close to one. Thus, one may want to choose large $c$ in order to maximize the PageRank mass of IN+SCC. However, in the next section we will indicate important drawbacks of this choice.

\section{PageRank mass of ESCC}
\label{sec:bounds}

Let us now consider the PageRank mass of the Extended SCC component
(ESCC) described in Section~\ref{sec:ergodic}, as a function of $c
\in [0,1]$. Subdividing the PageRank vector in the blocks
$\pi=[\pi_{\pureout}\;\pi_\escc]$, from (\ref{eq:vector_eq}) we
obtain
\begin{align}
\label{eq:piescc_sum}
||\pi_\escc(c)||&=(1-c)\gamma\ue[I-cT]^{-1}\onesccn,\end{align}
where $T$ represents the transition probabilitites inside the ESCC
block, $\gamma=|ESCC|/n$, and $\ue$ is a uniform probability
row-vector over ESCC. Clearly, we have $||\pi_\escc(0)||=\gamma$ and
$||\pi_\escc(1)||=0$. Furthermore, by taking derivatives we easily
show that $||\pi_\escc(c)||$ is a concave decreasing function. In
the next proposition (proved in the Appendix), we derive a series of
bounds for $||\pi_\escc(c)||$.

\begin{proposition}
\label{prop:bounds}
Let $\lambda_1$ be the Perron-Frobenius eigenvalue of $T$, and let
$p_1=\ue T\onesccn$
be the probability that the random walk started from a randomly chosen state in ESCC, stays in ESCC for one step.
\begin{itemize}
\item[(i)]
If $p_1<\lambda_1$ then
\begin{equation}
\label{ub} ||\pi_\escc(c)||<\frac{\gamma(1-c)}{1-c\lambda_1},\quad
c\in(0,1).\end{equation} \item[(ii)] If
$1/(1-p_1)<\ue[I-T]^{-1}\onesccn$ then
\begin{equation}
\label{lb} ||\pi_\escc(c)||>\frac{\gamma(1-c)}{1-cp_1},\quad
c\in(0,1).\end{equation}
\end{itemize}
\end{proposition}

The condition $p_1<\lambda_1$
 has a clear intuitive interpretation. Let $\hat{\pi}_\escc$ be the probability-normed
left Perron-Frobenius eigenvector of $T$. Then $\hat{\pi}_\escc$,
also known as a {\it quasi-stationary} distribution of $T$, is the
limiting probability distribution of the Markov chain given that
the random walk never leaves the block $T$ (see e.g.
~\cite{Seneta}). Since $\hat{\pi}_\escc T=\lambda_1$, the
condition $p_1<\lambda_1$ means that the chance to stay in ESCC
for one step in the quasi-stationary regime is higher than
starting from the uniform distribution $\ue$. Although
$p_1<\lambda_1$ does not hold in general, one may expect that it
should hold for transition matrices describing large entangled
graphs since quasi-stationary distribution should favor states,
from which the chance to leave ESCC is lower.

\begin{figure}[hbt]
              \centering {\epsfxsize=2in \epsfbox{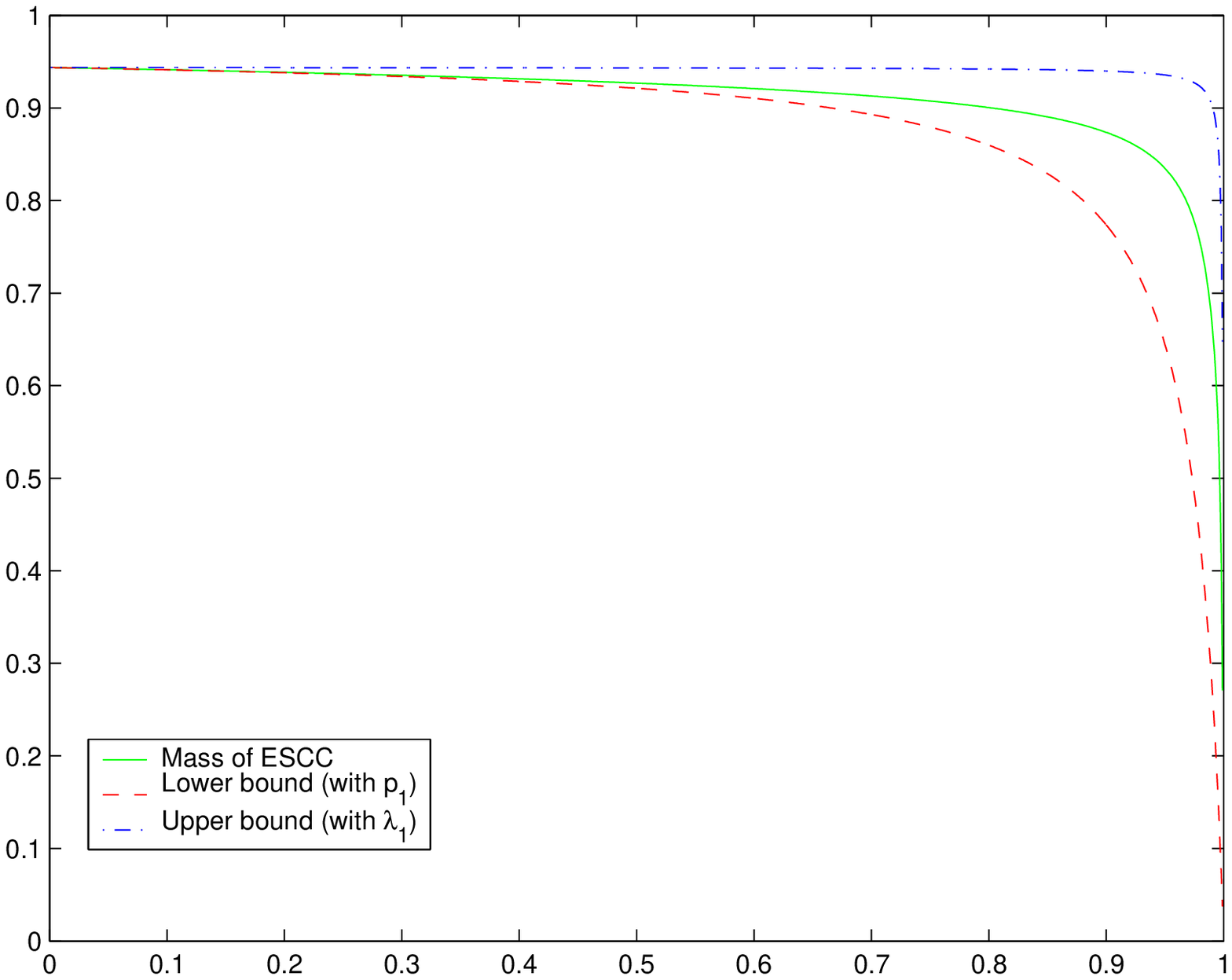}}
{\epsfxsize=2in \epsfbox{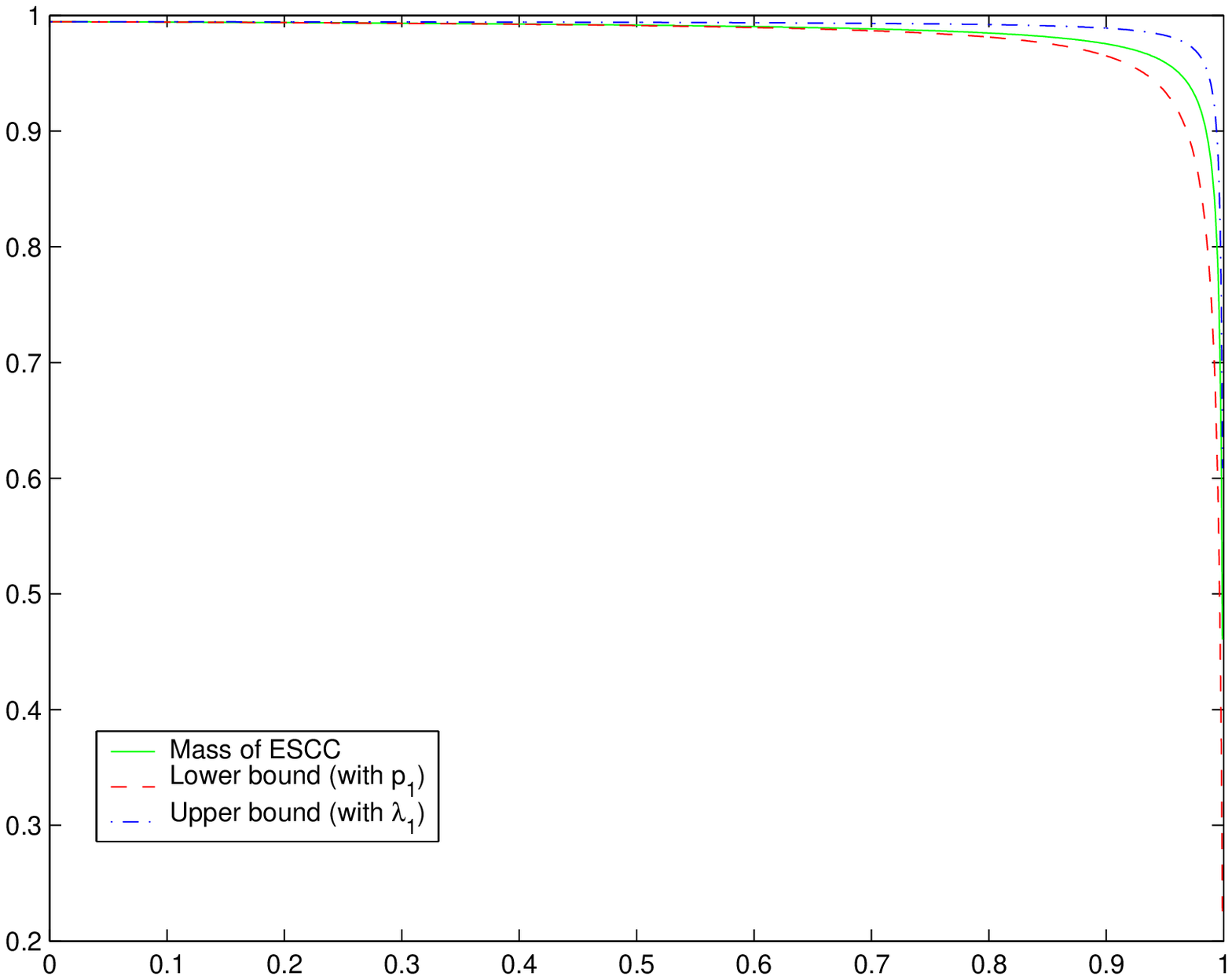}}
                \caption{\small PageRank mass of ESCC and bounds, INRIA (left) and FrMathInfo (right)}
\label{fig:escc}
\end{figure}

Both conditions of Proposition~\ref{prop:bounds} are satisfied in
our experiments. With the help of the derived bounds we conclude
that $||\pi_\escc(c)||$  decreases very slowly for small and
moderate values of $c$, and it decreases extremely fast when $c$
becomes close to 1. This typical behavior is clearly seen in
Figure~\ref{fig:escc}, where $||\pi_\escc(c)||$ is plotted with a
solid line.
The bounds are plotted in Figure~\ref{fig:escc} with dashed lines.
For the INRIA dataset we have $p_1=0.97557$, $\lambda_1=0.99954$,
and for the FrMathInfo dataset we have $p_1=0.99659$,
$\lambda_1=0.99937$.

From the above we conclude that the PageRank mass of  ESCC is
smaller than $\gamma$ for any value $c>0$. On contrary, the PageRank
mass of Pure OUT increases in $c$ beyond its ``fair share''
$\delta=|Pure OUT|/n$. With $c=0.85$, the PageRank mass of the Pure
OUT component in the INRIA dataset is equal to $1.95\delta$. In the
FrMathInfo dataset, the unfairness is even more pronounced: the
PageRank mass of the Pure OUT component is equal to $3.44\delta$.
This gives users an incentive to create dead-ends: groups of pages
that link only to each other. Clearly, this can be mitigated by
choosing a smaller damping factor. Below we propose one way to
determine an ``optimal'' value of $c$.

Let $\v$ be some probability vector over ESCC. We would like to
choose $c=c^*$ that satisfies the condition
\begin{equation}
\label{eq:c*} ||\pi_\escc(c)||=||\v T||,\end{equation} that is,
starting from $\v$, the probability mass preserved in ESCC after one
step should be equal to the PageRank of ESCC. One can think for
instance of the following three reasonable choices of $\v$: 1)
$\hat{\pi}_T$, the quasi-stationary distribution  of $T$, 2) the
uniform vector $\ue$, and 3) the normalized PageRank vector
$\pi_\escc(c)/||\pi_\escc(c)||$. The first choice reflects the
proximity of $T$  to a stochastic matrix. The second choice is
inspired by definition of PageRank (restart from uniform
distribution), and the third choice combines both these features.

If conditions of Proposition~\ref{prop:bounds} are satisfied, then
(\ref{ub}) and (\ref{lb})  hold, and thus the value of $c^*$
satisfying (\ref{eq:c*}) must be in the interval $(c_1,c_2)$,
where
\[(1-c_1)/(1-p_1 c_1)=||\v T||,\quad (1-c_2)/(1-\lambda_1 c_2)=||\v T||.\]
Numerical results for all three choices of $\v$ are presented in
Table~\ref{tab:c}.
\begin{table}[htb]
\centerline{
\begin{tabular}{|c|c|r|r|}
\hline
$v$&$c$&INRIA&FrMathInfo\\
\hline \hline
$\hat{\pi}_\escc$&$c_1$&0.0184&0.1956\\
&$c_2$&0.5001&0.5002\\
&$c^*$&.02&.16\\
\hline
$\ue$&$c_1$&0.5062&0.5009\\
&$c_2$&0.9820&0.8051\\
&$c^*$&.604&.535\\
\hline
$\pi_\escc/||\pi_\escc||$&$1/(1+\lambda_1)$&0.5001&0.5002\\
&$1/(1+p_1)$&0.5062&0.5009\\
\hline
\end{tabular}}
\vspace{.2cm} \caption{Values of $c^*$ with bounds.} \label{tab:c}
\end{table}

If $\v=\hat{\pi}_\escc$ then we have $||\v T||=\lambda_1$, which
implies $c_1=(1-\lambda_1)/(1-\lambda_1p_1)$ and
$c_2=1/(\lambda_1+1)$. In this case, the upper bound $c_2$ is only
slightly larger than $1/2$ and $c^*$ is close to zero in our data
sets (see Tabel~\ref{tab:c}). Such small $c$ however leads to
ranking that takes into account only local information about the Web
graph (see e.g. \cite{Fortunato06gn}). The choice
$\v=\hat{\pi}_\escc$ does not seem to represent the dynamics of the
system; probably because the ``easily bored surfer'' random walk
that is used in PageRank computations never follows a
quasi-stationary distribution since it often restarts itself from
the uniform probability vector.

For the uniform vector $\v=\ue$, we have $||\v T||=p_1$, which
gives $c_1,c_2,c^*$ presented in Table~\ref{tab:c}. We have obtained a higher upper bound but the
values of $c^*$ are still much smaller than $0.85$.

Finally, for the normalized PageRank vector
$\v=\pi_\escc/||\pi_\escc||$, using (\ref{eq:piescc_sum}), we
rewrite (\ref{eq:c*}) as
\begin{align*}
||\pi_\escc(c)||&=
\frac{\gamma}{||\pi_\escc(c)||}\pi_\escc(c)T\onesccn=
\frac{\gamma^2(1-c)}{||\pi_\escc(c)||}{\u}[I-cT]^{-1}T\onesccn,
\end{align*}
Multiplying by $||\pi_\escc(c)||$, after some algebra we obtain
\begin{eqnarray*}
||\pi_\escc(c)||^2&=
\frac{\gamma}{c}\,||\pi_\escc(c)||-\frac{(1-c)\gamma^2}{c}.
\end{eqnarray*}
Solving the quadratic equation for $||\pi_\escc(c)||$, we get
\[||\pi_\escc(c)||=r(c)=\left\{\begin{array}{ll}\gamma&\mbox{if }c\le 1/2,\\
\frac{\gamma(1-c)}{c}&\mbox{if }c>1/2.\end{array}\right.\] Hence,
the value $c^*$ solving (\ref{eq:c*}) corresponds to the point
where the graphs of $||\pi_\escc(c)||$ and $r(c)$ cross each
other. There is only one such point on (0,1), and since
$||\pi_\escc(c)||$ decreases very slowly unless $c$ is close to
one, whereas $r(c)$ decreases relatively fast for $c>1/2$, we
expect that $c^*$ is only slightly larger than $1/2$. Under
conditions of Proposition~\ref{prop:bounds}, $r(c)$ first crosses
the line $\gamma(1-c)/(1-\lambda_1c)$, then $||\pi_{T}(c)||_1$,
and then $\gamma(1-c)/(1-p_1c)$. Thus, we yield
$(1+\lambda_1)^{-1}<c^*< (1+p_1)^{-1}$. Since both $\lambda_1$ and
$p_1$ are large, this suggests that $c$ should be chosen
around 1/2. This is also reflected in Tabel~\ref{tab:c}.

Last but not least, to support our theoretical argument about the
undeserved high ranking of pages from Pure OUT, we carry out the
following experiment. In the $INRIA$ dataset we have chosen an
absorbing component in Pure OUT consisting just of two nodes. We
have added an artificial link from one of these nodes to a node in
the Giant SCC and recomputed the PageRank. In Table~\ref{tab:click}
in the column ``PR rank w/o link'' we give a ranking of a page
according to the PageRank value computed before the addition of the
artificial link and in the column ``PR rank with link'' we give a
ranking of a page according to the PageRank value computed after the
addition of the artificial link. We have also analyzed the log file
of the site INRIA Sophia Antipolis ({\tt www-sop.inria.fr}) and
ranked the pages according to the number of clicks for the period of
one year up to May 2007. We note that since we have the access only
to the log file of the INRIA Sophia Antipolis site, we use the
PageRank ranking also only for the pages from the INRIA Sophia
Antipolis site. For instance, for $c=0.85$, the ranking of Page A
without an artificial link is 731 (this means that 731 pages are
ranked better than Page A among the pages of INRIA Sophia
Antipolis). However, its ranking according to the number of clicks
is much lower, 2588. This confirms our conjecture that the nodes in
Pure OUT obtain unjustifiably high ranking. Next we note that the
addition of an artificial link significantly diminishes the ranking.
In fact, it brings it close to the ranking provided by the number of
clicks. Finally, we draw the attention of the reader to the fact
that choosing $c=1/2$ also significantly reduces the gap between the
ranking by PageRank and the ranking by the number of clicks.

\begin{table}[htb]
\centerline{
\begin{tabular}{|l|l|l|l|}
\hline
$c$&PR rank w/o link&PR rank with link&rank by no. of clicks\\
\hline \hline Node A & & & \\ \hline
0.5&1648&2307&2588\\
0.85&731&2101&2588\\
0.95&226&2116&2588\\
\hline Node B & & & \\ \hline
0.5&1648&4009&3649\\
0.85&731&3279&3649\\
0.95&226&3563&3649\\
\hline
\end{tabular}}
\vspace{.2cm} \caption{Comparison between PR and click based
rankings.} \label{tab:click}
\end{table}

To summarize, our results indicate that with $c=0.85$, the Pure OUT
component receives an unfairly large share of the PageRank mass.
Remarkably, in order to satisfy any of the three intuitive criteria
of fairness presented above, the value of $c$ should be drastically
reduced. The experiment with the log files confirms the same. Of
course, a drastic reduction of $c$ also considerably accelerates the
computation of PageRank by numerical methods
\cite{Avrachenkov07SIAM,LangvilleMeyer,Berkhin05}.

\section*{Acknowledgments}

This work is supported by EGIDE ECO-NET grant no. 10191XC and by NWO Meervoud grant no.~632.002.401.


\renewcommand{\theequation}{A.\arabic{equation}}
  \setcounter{equation}{0}  
\renewcommand{\thesubsection}{A.\arabic{subsection}}
  \setcounter{subsection}{0}  
  \renewcommand{\thelemma}{A.\arabic{lemma}}
  \setcounter{lemma}{0}  
\section*{Appendix}

\subsection{Results from Singular Perturbation Theory}

\begin{lemma}
\label{lm:sp} Let $A(\eps)=A-\eps C$ be a perturbation of
irreducible stochastic matrix $A$ such that $A(\eps)$ is
substochastic. Then, for sufficiently small $\eps$ the following
Laurent series expansion holds
\[[I-A(\eps)]^{-1}=\frac{1}{\eps}X_{-1}+X_0+\eps
X_1+... \ ,
\]
with
$$
X_{-1}=\frac{1}{\mu C \one} \one \mu,
$$
where $\mu$ is the stationary distribution of $A$. It follows that
\begin{equation}
\label{eq:LaurentX} [I-A(\eps)]^{-1}=\frac{1}{\mu \eps C \one}
\one \mu + \mbox{O}(1)\quad\mbox{as } \eps\to 0.
\end{equation}

\end{lemma}

\begin{lemma}
\label{lm:SPMC}
Let $A(\eps)=A+\eps C$ be a transition matrix of perturbed Markov chain.

The perturbed Markov chain is assumed to be ergodic for sufficiently small
$\eps$ different from zero. Let the unperturbed Markov chain $(\eps =0)$
have $m$ ergodic classes. Namely, the transition matrix $A$ can be written
in the form

$$
A=\left[ \begin{array}{cccc}
A_1 &        & 0   & 0 \\
    & \ddots &     &   \\
0   &        & A_m & 0 \\
L_1 & \cdots & L_m & E
\end{array} \right] \in \mathbb{R}^{n\times n}.
$$
Then, the stationary distribution of the perturbed Markov chain has a limit
$$
\lim_{\eps \to 0} \pi(\eps) =[\nu_1\mu_1 \ \cdots \ \nu_m\mu_m \ 0],
$$
where zeros correspond to the set of transient states in the unperturbed
Markov chain, $\mu_i$ is a stationary distribution of the unperturbed Markov
chain corresponding to the $i$-th ergodic set, and $\nu_i$ is the $i$-th
element of the aggregated stationary distribution vector that can be found
by solution
$$
\nu D = \nu, \quad \nu \one =1,
$$
where $D=MCB$ is the generator of the aggregated Markov chain and
$$
M=\left[ \begin{array}{cccc}
\mu_1 &        & 0   & 0 \\
    & \ddots &     &   \\
0   &        & \mu_m & 0
\end{array} \right] \in \mathbb{R}^{m\times n},\quad
B=
\left[ \begin{array}{cccc}
\one &        & 0    \\
    & \ddots &      \\
0   &        & \one  \\
\phi_1 & \cdots & \phi_m
\end{array} \right] \in \mathbb{R}^{n\times m}.
$$
with $\phi_i=[I-E]^{-1}L_i\one$.
\end{lemma}
The proof of this lemma can be found in
\cite{Avrachenkov99thesis,KorolyukTurbin,YinZhang}.

\subsection{Proofs}
\label{app:proofs}

{\it Derivation of (\ref{eq:piscc}).} First, we
observe that if $\piscc$ and $\pidn\onedn$ are known then it is
straightforward to calculate $\piout$. Namely, we have
$$
\piout = \piscc cR[I-cQ]^{-1} +\left(\frac{1-c}{n}+\pidn\onedn
\frac{c}{n}\right)\oneout^T[I-cQ]^{-1}.
$$
Therefore, let us solve the equations (\ref{eq:scc}) and
(\ref{eq:dn}). Towards this goal, we sum the elements of the
vector equation (\ref{eq:dn}), which corresponds to the
postmultiplication of equation (\ref{eq:dn}) by vector $\onedn$.
$$
-\piscc cS\onedn+\pidn\onedn-\frac{c}{n}\pidn\onedn\onedn^T\onedn
=\frac{1-c}{n}\onedn^T\onedn
$$
Now, denote by $n_{OUT}$, $n_{SCC}$
and $n_{DN}$ the number of pages in OUT component, SCC component
and the number of dangling nodes. Since $\onedn^T\onedn=n_{DN}$, we have
$$
\pidn\onedn=\frac{n}{n-cn_{DN}}(\piscc
cS\onedn+\frac{1-c}{n}n_{DN}).
$$
Substituting the above expression for $\pidn\onedn$ into
(\ref{eq:scc}), we get
$$
\piscc\left[I-cP-\frac{c^2}{n-cn_{DN}}S\onedn\onescc^T\right]=
\frac{c}{n-cn_{DN}}\frac{1-c}{n}n_{DN}\onescc^T+\frac{1-c}{n}\onescc^T,
$$
which directly implies (\ref{eq:piscc}).

{\it Proof of Proposition~\ref{prop:pure_out}} First, we note that if we make a change of
variables $\eps = 1-c$ the Google matrix becomes a transition
matrix of a singularly perturbed Markov chain as in
Lemma~\ref{lm:SPMC} with $C=\frac{1}{n}\one \one^T -P$. Let us
calculate the aggregated generator matrix $D$:
$$
D=MCQ=\frac{1}{n}\one\one^TQ-MPQ.
$$
Using $MP=M$, $MQ=I$, and $M\one=\one$ where vectors $\one$ are of
appropriate dimensions, we obtain
\begin{align*}
D&=\frac{1}{n}\one\one^TQ-I=
\frac{1}{n}\one[n_1+\one[I-{T}]^{-1}{R}_1\one,\cdots,
n_m+\one[I-\tilde{T}]^{-1}{R}_m\one]-I,
\end{align*}
where $n_i$ be the number of nodes in the block $Q_i$,
$i=1,\ldots,m$. Since the aggregated transition matrix $D+I$ has
identical rows, its stationary distribution $\nu$ is just equal to
these rows. Thus, invoking Lemma~\ref{lm:SPMC} we obtain
(\ref{eq:barpi}). 

\medskip

{\it Proof of Proposition~\ref{prop:pitilda}} Multiplying both sides of (\ref{eq:pitilda}) by $\onescc$
and taking the derivatives, after some tedious algebra we obtain
\begin{align}
||\tilde{\pi}'_\scc(c)|| \label{eq:mainterm'} =&
-a(c)+\frac{\beta}{1-c\beta}\,||\tilde{\pi}_{scc}(c)||,
\end{align}
where the real-valued function $a(c)$ is given by
\[a(c)=\frac{\alpha}{1-c\beta}\u[I-cP]^{-1}[I-P][I-cP]^{-1}\onescc.\]
Differentiating (\ref{eq:mainterm'}) and substituting
$\frac{\beta}{1-c\beta}||\tilde{\pi}_{SCC}(c)||$ from
(\ref{eq:mainterm'}) in the resulting expression, we get
\begin{align*}
||\tilde{\pi}''_\scc(c)||
&=\left\{-a'(c)+
\frac{\beta}{1-c\beta}a(c)\right\}+\frac{2\beta}{1-c\beta}\,||\tilde{\pi}'_{SCC}(c)||.
\end{align*}
Note that the term in the curly braces is negative by definition
of $a(c)$. Hence, if $||\tilde{\pi}'_\scc(c)||\le 0$ for some
$c\in[0,1]$ then $||\tilde{\pi}''_\scc(c)||<0$ for this value of
$c$.

\medskip

{\it Proof of Proposition~\ref{prop:bounds}} (i) The function $f(c)=\gamma(1-c)/(1-\lambda_1 c)$
is decreasing and concave, and so is $||\pi_\escc(c)||$. Also,
$||\pi_\escc(0)||=f(0)=\gamma$, and $||\pi_\escc(1)||=f(1)=0$.
Thus, for $c\in (0,1)$, the plot of $||\pi_\escc(c)||$ is either
entirely above or entirely below $f(c)$. In particular, if the
first derivatives satisfy $||\pi'_\escc(0)||<f'(0)$, then
$||\pi_\escc(c)||<f(c)$ for any $c\in(0,1)$. Since
$f'(0)=\gamma(\lambda_1-1)$ and $||\pi'_\escc(0)||=\gamma(p_1-1)$,
we see that $p_1<\lambda_1$ implies (\ref{ub}).

The proof of (ii) is similar. We consider a concave decreasing
function $g(c)=\gamma(1-c)/(1-p_1 c)$ and note that $g(0)=\gamma$,
$g(1)=0$. Now, if the condition in (ii) holds then
$g'(1)>||\pi'_\escc(1)||$, which implies (\ref{lb}).

\end{document}